\documentstyle[12pt]{article}
\begin{document}
\newcommand{\be}{\begin{equation}}
\newcommand{\ee}{\end{equation}}
\vskip 2cm
\begin{center}
{\large{\bf Time and Energy in Gravity Theory \\}}
\vspace*{1cm}  Ivanhoe B. Pestov\\
\vspace*{1cm} Bogoliubov  Laboratory of Theoretical Physics, Joint Institute
for Nuclear Research,\\  141980 Dubna, Moscow Region, Russia \end{center}

 \begin{abstract}
A new concept of internal time (viewed as a scalar temporal field)
is introduced which allows one to solve the energy problem in General
Relativity. The law of energy conservation means that the total energy
density of the full system of interacting fields (including gravitational
field) does not vary with time, thus being the first integral of the
system. It is demonstrated that direct
introduction of the temporal field permits to derive the general covariant
four dimensional Maxwell equations for the electric and magnetic fields
from the equations of electromagnetic fields considering in General
Relativity. It means that the fundamental physical laws are in full
correspondence with the essence of time. Theory of time presented
here predicts the existence of matter outside the time.
\vskip 0.5cm

\end{abstract}

\section{Introduction}
In the theory of gravitational field the problems
connected with  the energy conservation exist in a literal sense since
the time of its creation when
Einstein set up the problem of including gravity into the framework of the
Faraday concept of field. Thorough and deep
analysis of the problem of gravity field given by him in the works [1],
enables to formulate the key principles of gravity physics (General
Relativity). However, till now in the framework of these principles there
is no adequate solution to the energy conservation problem [2-5]. In
presented paper we give the simple solution of the problem in question,
which is based on the connection between the time and energy and
necessarily follows from the first principles of General Relativity if one
puts them into definite logical sequence. The energy conservation means
that total energy density of gravity field and all other fields does not
vary with time and hence represents the first integral of the system.

New understanding of time presented here may have
implications for the problem of time in quantum gravity. In fact,
a major conceptual problem in this field is the notion of time  and
how it should be treated in the formalism. The importance of this issue
was recognized at the beginning of the history of quantum gravity [6], but
the problem is still unresolved and has drawn recently an increasing
attention (see, for example [7],[8] and [9] and further references therein).

The paper is organized as follows. In section 2 we formulate
the fundamentals of general covariant theory of time with the guiding
idea that the  manifold is the main notion in physics and that time itself
is a scalar field on the manifold which defines the evolution of the full
system of fields being one of them. The system of interacting fields is
considered to be full if the gravitational field is included in it.

All known dynamical laws of nature have the following form:
the rate of
change of certain quantity with time equals to the results of action
of some operator on this quantity. So, a general covariant
definition of rate of change of any field with time
is one of the main results of the theory of time presented here. It is
also a starting point and relevant condition for the consideration of the
concept of evolution and the problem how to write the field equations in
the general covariant evolution form.

In section 3 the connection of the temporal field with
the Einstein gravitational potentials is established. This gives
the possibility to derive general covariant law of energy conservation
and produce the general covariant expression for the
energy density of the gravity field.

To demonstrate a concrete application of
general theory, in section 4 it is shown how evolution equations for
the vectors of electric and magnetic fields in the four dimensional
general covariant form can be derived from the equations for the
bivector of electromagnetic field.
Through this it is shown that the original Maxwell
equations (which as a matter of fact express the fundamental physical law)
are in full correspondence with fundamentals of the theory of time. In
section 5  different representations for the energy density of
the gravity field  are derived .

\section{Central role of time in gravity theory}

In accordance with the principles of gravity physics, in this section
the fundamentals of the theory of time are formulated.

According to Einstein, in presence of gravitational field all the systems of
coordinates  are on equal footing and in general, coordinates have
neither physical nor geometrical meaning. Thus, in the gravity
theory the coordinates play the same role as the Gauss coordinates in his
internal geometry of surfaces from which all buildings of the modern
geometry are grown.
Hence, one needs to construct the internal theory of physical fields
analogous to the Gauss internal geometry of surfaces. In other words, the
problem is how the modern differential geometry transforms into the
physical geometry.

The fields characterize the events and fill in the geometrical space.
In view of what has been said above, this space is a four dimensional
smooth manifold because this structure does not distinguish intrinsically
between different coordinate systems (the principle of general covariance
is naturally included into this notion).  Hence it follows that the notion
of smooth manifold is primary one not only in differential geometry but in
the theoretical physics as well that deal with gravitational phenomena.
This means that all other definitions, notions and laws should be
introduced into the theory through the notion of smooth manifold. Indeed
if some notion, definition or law is in agreement with the structure of
smooth manifold, then they are general covariant, i.e., do not depend on
the choice of coordinate system.

Definition of manifold is considered to be known and we only notice
that  all information on this topic can be found for example in reference
[4] or [5]. For our purposes it is enough to keep in mind that all smooth
manifolds can be realized as surfaces in the Euclidean space.
In what follows we shall consider only four dimensional manifolds. That is
evident from the physical point of view. However there is also a deep purely
mathematical reason for this choice.
Smooth manifold consists of topological manifold and
differential structure defined on it.
It is known [10], that a topological manifold always admits differential
structure if and only if its dimension is not larger than four.

It is clear that, in general, manifold should be in some relation
with its material content i.e., the fields. As it comes out,
the relation between them can be viewed as a form and content.
In view of this it is very important to know how
material content designs manifold. If we take the point of view that
manifold apriori is arena for the physical events, then it
is natural to select simplest manifold, for example, manifold of special
theory of relativity. However, Einstein argued (in the papers
mentioned above) that nature of gravity field is not compatible with  an
apriori defined manifold.

It can be shown that manifold as a surface in the Euclidean space is
designed by the covariant symmetrical tensor field  $g_{ij}$  on the
manifold, for which  adjoined quadratic differential form (Riemann metric)
\begin{equation}
ds^2 = g_{ij} du^i du^j,
\end{equation}
is positive definite.
Thus, covariant positive definite symmetrical tensor field
$g_{ij}(u)$ is the necessary element of any  general covariant
intrinsically self-consistent physical theory. Here we only give the
defining system of differential equations $$g_{ij}(u^1, \, u^2, \, u^3,
	     \, u^4) = \delta_{ab} \frac{\partial F^a}{\partial u^i}
\frac{\partial F^b}{\partial u^j}, \quad a,b = 1, \cdots ,4+k, \, k \geq 0
  $$ avoiding detailed consideration of this point.  If the functions
$g_{ij}(u^1,\, u^2, \, u^3, \, u^4)$ are known in local system of
coordinates $u^1, u^2, u^3, u^4$, then solving this system of equations we
obtain the functions $F^a(u^1, \, u^2, \, u^3, \, u^4)$ and hence the
region of manifold, defined by the equations $$x^a = F^a(u^1, \, u^2, \,
u^3, \, u^4), $$ where $x^a $ are the Cartesian coordinates of embedding
Euclidean space.

An important conclusion that follows from this consideration is that
there is one and only one way for the other fields to design manifold which
can be explained as follows. Let $u^1, u^2, u^3, u^4$  be a local
system of coordinates in the vicinity of some point of an abstract
manifold. Let us determine in such a vicinity the system of
differential equations that connect basis field $g_{ij}(u)$ with other
ones. Solving this system of equations we find field $g_{ij}(u)$ and by
doing so, we design a local manifold of physical system in question. In what
follows, a smooth manifold that corresponds to a physical system will be
called a physical manifold.

For further consideration of the principles of internal field theory
we simply note that there is a fundamental difference
between physics and geometry. In geometry there is no motion
that is tightly connected with the concept of time.
Thus, to be logical, we need to introduce time into the theory using
its first principles. Since, in general,
coordinates have no physical sense, time can be presented as a set of
functions of four independent variables (or in more strict manner as a
geometrical object on the manifold). It is quite obvious from the logical
point of view.

We put forward the idea that the time is a scalar field on
the manifold. By this we get a simple answer to the
question with long standing history "What is time ?" It should be
emphasized, that temporal field (together with other fields) designs
manifold as it was explained above but it has also another functions which
will be considered below.

Temporal field with respect to the coordinate system
$ u^1, u^2, u^3, u^4 $ in the region $U$ of smooth four dimensional manifold
$M$  is denoted as $$f(u) = f(u^1, u^2, u^3, u^4).$$ If the temporal field
is known, then to any two points $p$ and $q$ of manifold one can put in
correspondence an interval of time \begin{equation} t_{pq} = f(q) - f(p) =
\int \limits_{p}^{q} \partial_i f du^i.  \end{equation}

Unlike time, space is not an independent entity.
Instead of space we shall consider space cross-sections of the  manifold
$f^{-1} (t),$ which are defined by the temporal field.  For
the real number $t,$  space cross-section is defined by the equation
\begin{equation}
f(u^1, u^2, u^3, u^4) = t.
\end{equation}
One can call the number $t$ "the height" of the space cross-section
of manifold. If a point $p$ belongs to the space cross-section $f^{-1}
(t_1),$ and a point $q$  to the space cross-section $f^{-1} (t_2),$ then
the time interval (2) is equal to the difference of the heights  $t_{pq} =
t_2 - t_1.$ It is clear that $t_{pq}=0 $ if $p$ and $q$ belong to the same
space cross-section. Thus, to the every  segment of curve one can put in
correspondence a time interval.

Given the general covariant definition of time, one should show
that it is constructive in all respects. First of all we consider how the
temporal field defines the form of physical laws. It is known that the
general form of physical laws is very simple and is based on the following
recipe:  the rate of change of a certain quantity with time is
equal to the result of action of some operator on this quantity. To be
concrete, let us consider Maxwell equations. We know that the rate of
change of electrical and magnetic fields enter the dynamical equations
of electromagnetic field. Thus, we need to give a general covariant
definition of the rate of change of any field with time and in particular
this definition should be applicable for the case of electromagnetic field.

This problem has fundamental meaning, since it is impossible to speak
about physics when one has no mathematically rigorous definition of
evolution. It is clear that correct general covariant definition should
be conjugated with simple condition: if the rate of change of some field
with time is equal to zero in one coordinate system, then in any other
coordinate system the result will be the same.

On the manifold there is only one general covariant operation that can be
considered as a basis for the definition of the rate of change with time
of any field quantity. This general covariant operation is called
derivative in given direction and is defined by the vector field on
the manifold and the structure of the manifold itself. Thus, the problem is
to connect the temporal field  $f(u^1, u^2, u^3, u^4)$ with some vector
field $t^i (u^1, u^2, u^3, u^4).$  Since a temporal field is a scalar
one, the partial derivatives  define covector field $t_i = \partial_i
f.$  Now, the following definition becomes self-evident:
the gradient of temporal field (or the stream of time) is the
vector field of the type
\be
t^i = (\nabla f)^i = g^{ij} \frac{\partial
f}{\partial u^j} = g^{ij} \partial_j f = g^{ij} t_j,
\ee
where $g^{ij}$ are the contravariant components of the Riemann metric (1),
which is defined as usual, $g^{ij} g_{kj} = \delta^i_k .$ The gradient of
the field of time defines the direction of the most rapid  increase
(decrease) of the field of time. We define now the rate of change of some
quantity  with time as the derivative in the direction of the gradient of
the field of time and denote this operation by the symbol $D_t$.

Let us find the expression for the rate of change with time of the
temporal field itself. We have,
$$D_t f =t^i \partial_i f = g^{ij}
\partial_i f \partial_j f. $$
Since  $ D_t f$ is a general covariant
generalization of the evident relation
$\frac{d}{dt} \, t = 1,$  the temporal field should obey the
fundamental equation
 \be
(\nabla f)^2 = g^{ij} \frac{\partial f}{\partial u^j}
\frac{\partial f}{\partial u^j} =1.
\ee
Equation (5) means that the rate of change of the temporal field with time
is a constant quantity, the most important constant of the theory. From the
geometrical point of view the equation (5) simply shows that the gradient
of the temporal field is unit vector field on the manifold with respect
to the scalar product that is defined as usual by the metric (1), $$ (V, W)
= g_{ij} V^i W^j = g^{ij} V_i W_j = V^i W_i = |V| |W| \cos \phi.$$

It should be noted that the equation $(\nabla f)^2 = 1 $ is the main
equation of the geometrical optics. In view of this one can consider
equation (5) as the equation of 4-optics. This analogy can be
useful for consideration of some special problems in the theory of
time.

The rate of change of the vector field and symmetrical tensor field
with time is respectively given by the expressions
\begin{equation}
D_tV^i = t^k \frac{\partial V^i}{\partial u^k}
- V^k \frac{\partial t^i}{\partial u^k}, \end{equation}

 \begin{equation} D_tg_{ij} =  t^k \frac{\partial
 g_{ij}}{\partial u^k} + g_{kj} \frac{\partial t^k}{\partial u^i} +
 g_{ik} \frac{\partial t^k}{\partial u^j}.
 \end{equation}

Similar formulas can be presented for any other geometrical quantities.
In mathematical literature the derivative with respect to the given
direction is usually called  Lie derivative.  Thus, one can say that the
rate of change of any field with time  is the Lie derivative with respect
to the direction of the stream of time.

Consider the notion of time reversal and the invariance with respect to this
symmetry that is very important for what follows. It is almost evident that
in general covariant form  the time reversal invariance   means that theory
is invariant with respect to the transformations
\be	t^i \rightarrow  -  t^i.  \ee
It is clear that theory will be time reversal invariant if the
gradient of temporal fields will appear in all formulae only as an even
number of times, like $t^i t^j.$

Within the scope of special relativity and quantum mechanics, time and
energy are tightly connected. It is natural to  suppose that in gravity
physics the link between time and energy  even more deep and energy
conservation follows from the invariance of the Lagrangian theory with
respect to the transformations \be f(u) \rightarrow f(u)+a , \ee where $a$
 is arbitrary constant.  This invariance means that all the space sections
of manifold of system in question are equivalent.

Einstein himself put in correspondence to the gravity field symmetrical
tensor field ${\tilde g}_{ij},$ which is characterized by the condition
that adjoined quadratic differential form
\begin{equation}
ds^2 ={\tilde g}_{ij} du^i du^j,
\end{equation}
has the signature of the interval in special relativity. In accordance with
the principle of gravity physics discussed above, it is natural to
assume that Einstein's interval (10) has a structure that is defined by
the form-generating field $g_{ij}$ (Riemann's metric (1)) and temporal
field. If disclosed, this structure will give a simple method to introduce
temporal field into the equations of gravitational physics. It is quite
evident that the metric (1) has an Euclidean signature and hence it has no
structure. To be transparent in our consideration, let us
give a simple mathematical construction. Let $(V,V) = g_{ij} V^iV^j =
|V|^2  $ be usual scalar product  that is defined by
the metric (1).  One can consider tensor field $S^i_j$ as linear
transformation ${\bar V}^i = S^i_j V^j $ in the vector space in question.
If the operator $S$ is self-adjoint, that is $(V, SW) = (SV, W),$ then it is
always possible to introduce the scalar product associated with operator
$S$ via the formula $<V,V> = (V, SV).$ It is clear that  associated scalar
product will be in general indefinite and with respect to the initial
scalar product it has a structure. Thus, in general, the connection between
the forms (1) and (10) is given by the relation \begin{equation} {\tilde
g}_{ij} = g_{ik}S^k_j.  \end{equation} We shall give now simple expression
for the operator  $S^i_j,$  which defines the Einstein interval.
To this end we shall consider T-symmetry or time reversal invariance
(8) as a fundamental physical principle.  Further, we shall define
$T$- symmetry in the space of the vector field in order to
have transformation (8) as a particular case. We say that vector
fields ${\tilde V}^i$ and $V^i $ are $T$- symmetrical, if the sum of
this fields  is orthogonal to the gradient of temporal field and their
difference is collinear to it,
$$ ({\tilde V}^i+ V^i) t_i =
0, \quad {\tilde V}^i - V^i = \lambda t^i.  $$
From this we find,
$${\tilde V}^i = V^i - 2 n^i (V ,  n) = (\delta^i_j - 2n^i n_j) V^j,$$
where
$$ n^i = \frac{t^i}{\sqrt{(t,t)}}, \quad (t,t) = g_{ij} t^i t^j.$$

From this formula it follows that the fields  $ t^i$ and $ -t^i$ are
$T$-symmetrical and hence the definition of $T$-symmetry given in the
space of the vector field is correct. It is not difficult to verify that
$$ ({\tilde V} , W) = (V, {\tilde W}), \quad ({\tilde V},
{\tilde V})= ( V, V).$$
From the above consideration it follows that operator $S$ is
$T$- operator  that is $$S^i_j = \delta^i_j - 2n^i n_j $$ and in accordance
with (11) for the Einstein's potentials  we obtain the following expression
\be {\tilde g}_{ij} = g_{ik} (\delta^k_j - 2n^k n_j) = g_{ij} - 2 n_i
n_j,\ee which is invariant under the transformations (8).  To the tensor
field $\tilde g_{ij}$ one can put in correspondence contravariant tensor
field $\tilde g^{ij} = g^{ij} - 2 n^i n^j,$  which obeys the relation $$
\tilde g^{ik} \tilde g_{jk} = \delta^i_j.$$

Let us give the physical meaning of the Einstein's scalar product associated
with $T$- symmetry. Since
$$ < V, V>= (V,V) - 2 (V,n)^2 =
|V|^2 (1- 2 \cos ^2 \phi) = - |V|^2 \cos2\phi ,$$
where $\phi$  is the
angle between the vectors $V^i$ and $n^i$, then the Einstein's scalar
product is indefinite and can be positive, negative or equal to zero
according to the value of the angle $\phi.$ In particular,
$ <V, V > =0,$ if $\phi = \pi/4.$ Thus, the Einstein's  $T$- symmetrical
scalar product permits to classify all the vectors depending on
which  angle they form with the gradient of the temporal field.
As we see, the temporal field and T-symmetry define the Einstein's form
(10) as the metric of the normal hyperbolic type.  Hence, the gradient
of the temporal field defines the causal structure on the physical
manifold and can be identified with it. It is the physical meaning of the
Einstein's interval.

Let us show that the causal structure (the gradient of the temporal
field) can be reduced to the canonical form $(0,0,0,1)$ by the suitable
coordinate transformation at once in all points of some (may be small)
patch of any point on the physical manifold.  Local coordinates with
respect to which gradient of the temporal field has the form $(0,0,0,1)$
will be called compatible with causal structure.

Geometrically the stream  of time is defined as a congruence of lines (lines
of time) on the manifold.  We recall that the congruence of lines is a set
of lines characterized by the fact that only one element of the set passes
through each point of manifold (or its open region).  According to the
definition, lines belonging to the congruence do not intersect and fill
either the whole manifold or a part of it. The simple non-trivial example is
the congruence of rays coming from one point of the Euclidean space.

Analytically the lines of time are defined as the solutions of the
autonomous system of differential equations
 \begin{equation} \frac{du^i}{dt} = \; g^{ij}
\frac{\partial f}{\partial u^j} = g^{ij} \partial_j f = (\nabla f)^i ,
\quad (i=1,2,3,4).  \end{equation}
It can be shown that if the
functions $\varphi^i(t)$ are solutions to Eqs.
(13), the functions $\psi^i(t) = \varphi^i(t+a),$ where  $a$ is constant,
will also be solutions to them. Since $ds^2
=g_{ij}du^i du^j,$  from Eqs. (5) and (13) it follows that
$ds/ dt = \pm 1.$ Thus, the length of the time line is a linear
function of the parameter $t, \quad s = \pm t + a$.  After these
preliminary remarks, it is time to give solution of the above set problem.

Let
\be
	u^i(t) = \varphi^i(u^1_0,\, u^2_0, \, u^3_0, \, u^4_0, t) =
\varphi^{i}(u_0,t) \ee
be the solution to equations (13) with initial data
$\varphi^{i}(u_0, t_0) = u^i_0 $ so that
\be \frac{\partial {\varphi^i(u_0, t_0)}}{\partial u^j_0} = \delta^i_j.  \ee
Substituting  $u^i(t) = \varphi^i(u_0, t)$ into the function
$f(u^1, \,u^2,\,u^3,\, u^4)$ we obtain
$p(t) = f(\varphi(u_0,t)).$ Differentiating this function with respect to
$t,$ by virtue of (5) and (13), one finds
$$
\frac{dp(t)}{dt} = \frac{\partial f}{\partial u^i} \frac{du^i}{dt}=
g^{ij}\frac{\partial f}{\partial u^i} \frac{\partial f}{\partial u^j}  =
1.$$
It leads to
\be f(\varphi(u_0,t)) = t - t_0 + f(u_0).
\ee
Suppose that all initial data belong to the space section
\be f(u^1_0,\, u^2_0, \, u^3_0, \, u^4_0 ) = t_0.  \ee
Rewriting Eq. (17) in the parametric form
$$u^i_0 = \psi^i(x^1, x^{2}, x^3) ,$$ Eqs.
(14) can be written as the system of relations
 \be u^i = \phi^i (x^1, x^{2}, x^3, t). \ee
The right hand side of Eqs.(18) has continuous partial derivatives with
respect to variable  $x^1, x^{2}, x^3, t.$ From (14) and (15) it follows
that the functional determinant of the system  (18) is not equal to
zero and hence it designes internal system of coordinate of the dynamical
system in question. Now one can show that in such a system of coordinates the
covariant and contravariant components of the gradient of the temporal field
and some components of the field $g_{ij}$ take a simple numerical
value $$ t^i =(0,0,0,1) = t_i, \quad
g_{44} = g^{44} =1, \quad g_{\mu4} = g^{\mu4} =0, (\mu =1,2,3)$$ and
hence this coordinate system is compatible with causal structure.
Indeed, since in the system of coordinates $x^1,
x^{2}, x^3, t$ temporal field has a simple form  $$f(x^1, x^2, x^3, t)=
f(\phi(x,t)) = t ,$$ then in this coordinate patch $$\frac{\partial
f}{\partial x^1} = \frac{\partial f}{\partial x^2} = \frac{\partial
f}{\partial x^{3}} = 0, \quad \frac{\partial f}{\partial t} = 1 $$ and
hence $t_i = (0,0,0,1).$ Since  coordinates  $x^1, x^2, x^{3} $ do not
vary along the lines of time, then $$ \frac{dx^{\mu}}{dt} =
g^{\mu 4} = 0, \quad (\mu = 1,2,3) \quad \frac{dt}{dt} = g^{44} = 1.$$
Thus, in the internal system of coordinates
$$g^{\mu 4} = 0, \quad g^{44} = 1, \quad t^i =g^{ij} t_j = g^{i4} =
(0,0,0,1),$$ which is what we had to prove. Therefore, it follows
that in the system of coordinates compatible with causal structure,
metric  (1) takes the form\be ds^2 = g_{\mu\nu} dx^{\mu}
dx^{\nu} + (dt)^2, \quad \mu,\nu =1,2, 3 ,\ee since $t_i = g_{ij} t^j =
g_{i4}.$

So, for any point of physical manifold we can indicate a local
coordinate neighborhood with the coordinates compatible with causal
structure. Covering manifold with such charts we get atlas on the
manifold that is compatible with its causal structure. In this atlas the
filed equations have the most simple form in the  sense that
all components of the gradient of temporal field   take numerical values.
It should be noted that the system of coordinates compatible with causal
structure is similar to the Darboux system of coordinate in the theory of
symplectic manifolds which is a geometrical basis for the Hamilton
mechanics .

From the above consideration it also follows that variable $t$,
parametrizing the line of time can be considered as
coordinate of time of the physical system in question. This name is
justified particularly by the fact that in accordance with (6) and (7)
the rate of change of any field with time is equal to the partial
derivative with respect to $t$, i.e., $D_t = \partial / \partial t $ in the
system of coordinate compatible with causal structure. This is due to
the fact that $f(x^1, x^2, x^3, t)= f(\phi(x,t)) = t .$

Furthermore, if we reverse time putting
${\tilde t}^i = - t^i,$ then the lines of time will be parametrized by
new variable  ${\tilde t.}$ From the equations (13) it is not
difficult to derive that there is one-to-one and
mutually continuous correspondence between the parameters  $t$ and ${\tilde
t}$  given by the relation
${\tilde t} = -t.$ From here it is clear that in the system of
coordinate defined by the time reversal, the variable $-t$ will be the
coordinate of time. Thus, the general covariant definition of time reversal
given by (8) is adjusted with the familiar definition that is connected with
the transformation of coordinates. Finally, we note one more important
fact related to the time coordinate. Transformations (9) take the
well-known form $t \rightarrow t + a $ in the system of coordinates
compatible with the causal structure.

\section{Equations of the gravity field}

Consideration made above gives the evident method of constructing
Lagrangians in gravity physics. It is clear
that dynamics of gravity field is defined by the Lagrangian
\be L = {\tilde R} ,\ee
where  ${\tilde R} $ is a scalar that is
constructed from the Einstein's gravitational potentials by the
standard way.

Since the Einstein interval is defined by the two fields connected by the
equation (5), it is necessary to pay special attention when deriving the
equations of gravitation field. A standard method is to incorporate the
constraint (5) via a Lagrange multiplier $\varepsilon = \varepsilon(u)$,
rewrite the action density  for gravity field in the form  $$ L_g = {\tilde
R} + \varepsilon(g^{ij} t_i t_j - 1)$$ and treat the components of the
fields $g_{ij}$ and $ f$  as independent variables.

Let us consider the action
\be A= \frac{1}{2} \int {\tilde R} \sqrt g \,d^{4}u + \int  L_F(\tilde g, q)
\sqrt g \,d^{4}u + \frac{1}{2} \int \varepsilon(g^{ij} t_i t_j - 1) \sqrt g
\, d^{4}u,\ee
where $g = Det(g_{ij}) > 0$ and $L_F(\tilde g, q) $
is the Lagrangian density of the system of other fields $q$
which incorporates the Einstein's gravitational potentials in the
conventional form. Such method of introduction of causal structure
into the equations of material fields does not require special explanation.
Nevertheless, it will be demonstrated on the example of the
derivation of the Maxwell equations for the electric and magnetic fields in
general covariant four-dimensional and evolutionary form since
this problem remains unresolved up-to-date.

Since Einstein's gravitational potentials are the functions of
$g_{ij}$ and  $f$, then it is necessary to use the rule of
differentiation of the complex function.
We have, $\delta \tilde R = {\tilde g}^{ij} \delta {\tilde R}_{ij} +
{\tilde R}_{ij} \delta {\tilde g}^{ij}. $ Further we denote the
Christoffel symbols as
 $ \Gamma^i_{jk}, \quad {\tilde \Gamma}^i_{jk}$
belonging to the fields $g_{ij},
\, {\tilde g}_{ij}$, respectively. From (12), after some calculations,
we obtain the formulae
\be{\tilde
\Gamma}^i_{jk} = \Gamma^i_{jk} + H^i_{jk}, \ee where \be H^i_{jk}= n^i
(\nabla_j n_k + \nabla_k n_j) + (2n^i n^l - g^{il}) (n_j \delta^m_k + n_k
\delta^m_j) (\nabla_m n_l - \nabla_l n_m) .  \ee In  (23)
$\nabla_j $  is covariant derivative with respect to the connection $
\Gamma^i_{jk}.$  It is easy to derive from (23) that $H^i_{ji} = 2 n^i
\nabla_j n_i = 0,$ since $n^i n_i = 1.$ Therefore, ${\tilde
\Gamma}^j_{jk} = \Gamma^j_{jk}, $ and hence $${\tilde \nabla}_l
V^l = \nabla_l  V^l = \frac{1}{\sqrt g} \partial_l (\sqrt g V^l),$$ where
${\tilde \nabla}_l $ is the covariant derivative with respect to
the connection $ {\tilde \Gamma}^i_{jk} .$ From this it follows that
${\tilde g}^{ij} \delta {\tilde R}_{ij}  $ can be omitted as a perfect
differential.  Varying now ${\tilde g}^{ij}, $ we get $$\delta {\tilde
g}^{ij} = \delta g^{ij} + P^{ij}_{kl}\delta g^{kl} ,$$ where $$P^{ij}_{kl}
= 2 n^i n^j n_k n_l - n^i (n_k \delta^j_l + n_l \delta^j_k) - n^j (n_k
\delta^i_l + n_l \delta^i_k) $$ is a tensor field symmetrical in covariant
and contravariant indices.  Thus, $$ \delta ({\tilde R} \sqrt g ) =
({\tilde R}_{ij} + {\tilde R}_{kl}P^{kl}_{ij} - \frac{1}{2}{\tilde R}
g_{ij}) \delta g^{ij} \sqrt g $$ with neglect of a perfect differential.
Let $ G_{ij} = {\tilde R}_{ij}- \frac{1}{2} {\tilde g}_{ij} {\tilde R}$ be
the Einstein's tensor.  Observing that ${\tilde g}_{ij} + {\tilde
g}_{kl}P^{kl}_{ij} = g_{ij},$  it is easy to verify that variation $\delta
({\tilde R} \sqrt g ) $  can be presented in the form $$ \delta ({\tilde R}
\sqrt g ) = ( G_{ij} + G_{kl} P^{kl}_{ij}) \delta g^{ij} \sqrt g.$$ Further
we put $\delta L_F = \frac{1}{2} M_{ij} \delta {\tilde g}^{ij}, $ and
introduce by  the standard way the energy-momentum tensor ${T}_{ij} =
M_{ij} - {\tilde g}_{ij}L_F.  $ Then $\delta (L_F \sqrt g) = \frac{1}{2}
(M_{ij} + M_{kl} P^{kl}_{ij} - g_{ij} L_F) \delta g^{ij}\sqrt g $ and total
variation of the action can be presented in the following form \be \delta A
= \frac{1}{2} \int ( G_{ij} + G_{kl} P^{kl}_{ij}+ {T}_{ij} +{T}_{kl}
P^{kl}_{ij} + \varepsilon t_i t_j ) \delta g^{ij}\sqrt g d^{4}u .  \ee
One can consider tensor $P^{kl}_{ij} $  as operator  $P$ acting in the
space of symmetrical tensor fields. The characteristic equation of
this operator has the form
$P^2 + 2P = 0 ,$ and hence $(P+1)^2 =1.$ Thus, operator $P+1$ is inverse to
itself. Since  $t_i t_j + t_k t_l
P^{kl}_{ij} = -t_i t_j , $ then it follows that Einstein equations have
the form
\be G_{ij} + {T}_{ij} = \varepsilon {\partial_i}f {\partial_j}f , \quad
g^{ij} {\partial_i}f {\partial_j}f =1.  \ee
The equations (25)  constitute the full system of
equations of gravity field. As it is shown above, these equations emerge
from the first principles of gravity physics formulated by Einstein.
Soon it will be shown that the Lagrange multiplier
$\varepsilon $ has a physical meaning of energy density of the system in
question. From the Eqs. (25) it follows that \be \varepsilon = G_{ij}t^i
t^j + {T}_{ij} t^i t^j .\ee

Varying the action with respect to $f$ and taking into account that
$\delta_f {\tilde g}^ {ij} = Q^{kij} \delta t_k,$ where  $$Q^{kij} =
\frac{2}{\sqrt (t,t)}(2n^k n^i n^j - g^{ki} n^j - g^{kj} n^i),$$ we
get the following equation \be \frac{1}{2} \nabla_k
\bigl( ( G_{ij} + {T}_{ij} - \varepsilon t_i t_j) Q^{kij} \bigr) + \nabla_k
(\varepsilon t^k ) = 0.  \ee From Eqs. (25) and (27) we have
 \be \nabla_k (\varepsilon t^k ) = 0.  \ee Equation
(28) expresses the law of energy conservation in gravitational
physics which, evidently, is general covariant.
To make sure that we indeed deal with conservation of energy, it is
sufficient to figure out that action (21) is invariant with
respect to transformation $f(u)
\rightarrow f(u) + a,$ where $a$ is constant and $f(u) $ is a temporal
field.  Thus, the equation (28) results from the Noether's theorem and
the invariance of the action with respect to the one parametric group
of transformations (9).

Consider the law of energy conservation from the various
points of view. The so-called  local energy conservation is written as
follows  $$ {\tilde \nabla}_i {T}^{ij} = 0,$$ where
${T}^{ij} = {T}_{kl} {\tilde g}^{ik} {\tilde g}^{jl}.$ These equations
are fulfilled on the equations of the fields $q$ that contribute to the
energy-momentum tensor. Since $$ {\tilde
\nabla}_i G^{ij} = 0$$ identically, then from the Einstein's
field equations (25) it follows that
$$ {\tilde \nabla}_i {T}^{ij} = \varepsilon t^i ({\tilde \nabla}_i t^{j}) +
t^j {\tilde \nabla}_i (\varepsilon t^i) .$$  We now show that $t^i
{\tilde \nabla}_i t^{j}  =0.$ From (5) we have $n_i = t_i,
\quad \nabla_i n_j - \nabla_j n_i  = \nabla_i t_j - \nabla_j t_i = 0 .  $
With this and from (22), (23) we get \be{\tilde
\Gamma}^i_{jk} = \Gamma^i_{jk} + 2 t^i \nabla_j t_k .  \ee Thus,
$t^i {\tilde \nabla}_i t^{j}  = t^i \nabla_i t^j + 2 t^it^j t^l \nabla_i
t_l  = 0.$ Since  ${\tilde \nabla}_i (\varepsilon t^i)= \nabla_i
(\varepsilon t^i), $  then finally we have $$
{\tilde \nabla}_i { T}^{ij} =  t^j \nabla_i (\varepsilon t^i) .$$  in
view of this the energy conservation law can be treated as the
condition of compatibility of the field equations. In this sense, the  law
of energy conservation is  analogous to the law of charge conservation.

Show that rate of change of the energy density
$D_t (\sqrt g \varepsilon) $ equal to zero and hence this quantity is a
first integral of the system in question.  We have
 $D_t (\sqrt g \varepsilon) = t^i \partial_i (\sqrt g \varepsilon) +
\sqrt g \varepsilon \partial_i  t^i = \partial_i (\sqrt g \varepsilon
t^i).$  Since $  {\sqrt g} \nabla_k (\varepsilon t^k ) = \partial_i
(\sqrt g \varepsilon t^i),$ then from the law of energy conservation (28)
it follows that energy density is the first integral of the system
 \be D_t (\sqrt g \varepsilon) = 0.  \ee
In the system coordinates, compatible with causal structure, this equation
has a more customary form $$
\frac{\partial}{\partial t} (\sqrt g \varepsilon) = 0. $$

Compare the law of energy conservation with the law of charge
conservation which can be written in general covariant form as
follows $\nabla_i J^i =0.$
Putting
$J^i = t^i (t,J) + J^i - t^i (t,J) = \rho t^i + P^i ,$  we find that
$D_t (\sqrt g \rho) + \partial_i (\sqrt g P^i) = 0.$  Thus, we see that the
charge density is not in general the first integral  of the system
while energy density always is.

\section{Time in the theory of electric and magnetic fields}

In this chapter the methods of gravitational physics and theory of time
are applied for the derivation of the general covariant  Maxwell equations
for the electric and magnetic fields.

In connection with this we remind that the process of
raising and lowering the indices is carried out with the
help of the metric tensor $g_{ij}$ and its reciprocal
$g^{ij}.$  Symbol $\nabla_i$ denotes the covariant derivative
with respect to the connection
$\Gamma^i_{jk}$ of the metric $g_{ij},
\quad\varepsilon_{ijkl}$ and $\varepsilon^{ijkl}$  are covariant and
contravariant components of the Levi-Chivita tensor normalized as
$ \varepsilon_{1234} = {\sqrt g},$ and $\varepsilon^{1234} = 1/ {\sqrt g},$
where $g$ is determinant of the metric tensor. Since metric is positive
definite then $g > 0.$

Let $A_i $ be the vector potential of the electromagnetic field.
Let us define the tensor of electromagnetic field as usual
\be F_{ij} = \partial_i A_j - \partial_j A_i.\ee  According to the
principle of gravity physics we introduce temporal field into the
theory of electromagnetic field through the gauge invariant and
general covariant Lagrangian \be
L_{em} = \frac{1}{4} F_{ij} F_{kl} {\tilde g}^{ik}  {\tilde g}^{jl}  .  \ee
From (32) we derive the energy-momentum tensor
\be T_{ij} = F_{ik} F_{jl} {\tilde g}^{kl}
- {\tilde g}_{ij} L_{em}.  \ee
Let us show that the equations for the
tensor of the electromagnetic field can be written in the form
\be  \nabla_i {\tilde F}^{ij} = 0 , \quad \nabla_i F^{ij} = 0,  \ee where
$$ {\tilde F}^{ij} = \frac{1}{2}\varepsilon^{ijkl} F_{kl}, \quad F^{ij}=
F_{kl} {\tilde g}^{ik} {\tilde g}^{jl}. $$
Indeed, from (31) we have
$$\partial_i F_{jk} + \partial_j F_{ki} + \partial_k F_{ij}=0.$$ In this
equation partial derivatives can be substituted by the covariant ones
and after contraction with the Levi-Chivita tensor  $\varepsilon^{ijkl}$
we get first of the equations (34). Further, from (29) it follows that
$$ {\tilde \nabla}_i F^{ij} = \nabla_i F^{ij} + H^i_{il} F^{lj} +
H^j_{il} F^{il} = \nabla_i F^{ij}, $$ Q.E.D.

Now we derive from the equations (34) the general covariant four
dimensional Maxwell equations for the electric and magnetic fields which
in fact express the fundamental physical laws. First of all we
formulate the main relations
of the vector algebra and vector analysis on the four dimensional physical
manifold which is interesting by itself.

A scalar product of two vector fields
$A^i$ and $B^i$ is defined as usual $(A,B) = g_{ij}
A^i B^j = A^i B_i = A_i B^i = g^{ij} A_i B_j =|A| |B| \cos \varphi .$
A vector product of two vector fields  $A^i$ and $B^i$ we shall
construct as follows   \be C^i = [A B]^i
=\varepsilon^{ijkl}t_j A_k B_l.  \ee It is evident that $[A B] +
[B A] = 0.$  From geometrical point of view a vector product
is a vector field that is tangent to any space section of the
manifold in any point. By the direct calculation it can be shown that
 $|[A B]| = |A| |B| \sin \varphi, \quad [A[B
C]] = B (A,C) - C(A,B).$ It should be noted that vector product is
invariant with respect to transformations
 $$ A^i \rightarrow A^i
+ \lambda t^i, \quad B^i \rightarrow B^i + \mu t^i,$$  where $\lambda$ and
$\mu$ are scalar fields.

Differential operators of the vector analysis on the physical manifold are
defined as natural as algebraic ones. For the divergence and gradient we
have respectively   \be  div \,A =   \nabla_i A^i  = \frac{1}{\sqrt g}
\partial_i (\sqrt g A^i),\ee \be ( grad \,\phi )^i = ( g^{ij}-t^i t^j)
\partial_j \phi = g^{ij} \partial_j \phi - t^i D_t \phi.  \ee From (36),
(37) we derive $$ div \, grad \phi = \frac{1}{\sqrt g} \partial_i (\sqrt g
g^{ij} \partial_j \phi) - \frac{1}{\sqrt g} \partial_i (\sqrt g  t^{i} D_t
\phi)= \nabla_i \nabla^i \phi - \nabla_i (t^i D_t \phi). $$

A rotor of the vector field $A$ is defined as a vector product of $ \partial
$ and $A$ \be ( rot\,A )^i = \varepsilon^{ijkl}t_j
\partial_k A_l = \frac{1}{2} \varepsilon^{ijkl}t_j ( \partial_k A_l -
\partial_l A_k ).\ee It is easy to verify that $$
rot \, grad \phi = 0.$$ Since the  temporal field satisfies the
equations $\partial_i
t_j -\partial_j t_i = 0, $ then rotor is invariant with respect to the
transformations $A^i \rightarrow A^i + \lambda t^i, $ where $ \lambda $ is a
scalar field.  It is evident that all operations so defined are
general covariant.

There is only one  direct way to derive from the eqs. (34) the
fundamental physical laws  first formulated by Maxwell. Consider
the rate of change with time  of the potentials of electromagnetic field.
By definition, we have $$ D_t A_i = t^k \partial_k A_i + A_k \partial_i t^k
= t^k (\partial_k A_i - \partial_i A_k) + \partial_i(t^k A_k) = t^k F_{ki}
- \partial_i \phi,$$  where $\phi = - t^k A_k. $ Thus, the rate of change
for the electromagnetic potentials can be presented as the difference of
two covector fields with one of them having the form \be E_i = t^k F_{ik}
.\ee From this it follows that strength of the electric field is general
covariant quantity that is defined by the equation
\be E_i = - D_t A_i - \partial_i \phi.  \ee
Now it is quite clear how one can give general covariant definition of the
magnetic field strength. According to  (38), the axial vector
field \be H^i = (rot A)^i = \varepsilon^{ijkl}t_j
\partial_k A_l = \frac{1}{2} \varepsilon^{ijkl}t_j ( \partial_k A_l -
\partial_l A_k ) \ee is the strength of the magnetic field. Out of (41)
it follows that \be H_i = t^k \tilde F_{ik}, \ee where $$\tilde
F_{ij} = g_{ik} g_{jl}\tilde F^{kl} .  $$

Given electric and magnetic fields we see that the tensor of the
electromagnetic field is simply a suitable notation.
It is evident that vectors $E$ and $H$  are orthogonal to gradient of
temporal field \be  t^i H_i =
t^i E_i = 0 .\ee
Thus, vectors $E^i$ and $H^i$ belong to the three dimensional
linear space of the vector fields orthogonal to the gradient of
the temporal field.

Now it is not difficult to derive the Maxwell equations for the electric
and magnetic fields staying in the framework of the vector analysis.
However, for some reasons it has a certain interest to give derivation of
the fundamental equations of physics precisely from the eqs. (34).

Resolving equations (39), (42) over  $F_{ik},$ we obtain \be
F_{ij} = - t_i E_j + t_j E_i - \varepsilon_{ijkl} t^k H^l .\ee Thus, on the
physical manifold there is general covariant one-to-one algebraic relation
between the electric and magnetic fields and tensor of the electromagnetic
field that is given by the equations
(39),(42), (43), (44).  Out of  (44) we find
\be {\tilde F}^{ij} = - t^i H^j + t^j H^i - \varepsilon^{ijkl} t_k E_l ,
\ee

\be
   F^{ij} =  t^i E^j - t^j E^i - \varepsilon^{ijkl} t_k H_l .
\ee

Substituting (45), (46) into (34), we shall obtain the
Maxwell equations for the strengths of the electric and magnetic
fields in the following general covariant form  \be D_t H^i + H^i \nabla_k
t^k = - \varepsilon^{ijkl} t_j \nabla_k E_l \ee
\be div \, H = \nabla_i
\,H^i = \frac{1}{\sqrt g} \partial_i(\sqrt g H^i) = 0 \ee
 \be D_t E^i + E^i
\nabla_k t^k = \varepsilon^{ijkl} t_j \nabla_k H_l \ee

 \be  div \, E  = \nabla_i \,E^i
= \frac{1}{\sqrt g} \partial_i(\sqrt g E^i) = 0. \ee

Derive, for example, first set of the Maxwell equations.
Substituting (45) into the first of Eqs.
(34), we get

$$ -\nabla_i {\tilde F}^{ij} = \nabla_i  ( t^i H^j - t^j H^i +
\varepsilon^{ijkl} t_k E_l) = $$
$$t^i \nabla_i H^j - H^i \nabla_i t^j + H^j \nabla_i t^i - t^j \nabla_i
H^i +\varepsilon^{jkil} t_k \nabla_i E_l. $$
Since, according to (6),
$$D_t H^j = t^i \partial_i H^j - H^i \partial_i t^j =
t^i \nabla_i H^j - H^i \nabla_i t^j ,                $$
we have
$$ -\nabla_i {\tilde F}^{ij} = D_t H^j +\varepsilon^{jkil} t_k \nabla_i E_l
+ H^j \nabla_i t^i - t^j \nabla_i H^i .  $$
Taking into account that
$$D_t \, t_i = D_t\, t^i = 0, \quad t_i H^i =0,$$
we derive from  the last equations  the equation (48) and then
equation (47).

Since $\sqrt g
\nabla_i t^i = \partial_i (\sqrt g t^i),$ $D_t \sqrt g = t^i \partial_i
\sqrt g + \sqrt g \partial_i t^i = \partial_i( \sqrt g t^i),$  then the
equations (47) and (49) may be written in more symmetrical form

$$ \frac{1}{\sqrt g}  D_t(\sqrt g  H^i)  = -
\varepsilon^{ijkl} t_j \nabla_k E_l ,$$

$$ \frac{1}{\sqrt g}  D_t(\sqrt g  E^i)  = \varepsilon^{ijkl} t_j
\nabla_k H_l $$

\be \frac{1}{\sqrt g}  D_t(\sqrt g  H)  = -
rot E \ee

\be \frac{1}{\sqrt g}
 D_t(\sqrt g  E)  = rot H .\ee

From Eqs. (51), (52) it follows that $$ D_t (\partial_i(\sqrt
g H^i)) = 0 , \quad D_t (\partial_i(\sqrt g E^i)) = 0 , $$ and
therefore,
Eqs. (48), (50) may be considered as the initial conditions.
From Eqs. (51), (52) one can also derive  that $$ D_t
(\sqrt g t_i E^i) = 0 , \quad D_t (\sqrt g  t_i H^i) = 0 , $$ which
shows that the orthogonality of electric and magnetic fields to the gradient
of the temporal field may be considered as the initial conditions.

Thus, it is shown that the principles of the gravity physics
are in full correspondence with the fundamental physical laws and hence
they can be considered as a method to derive  new fundamental equations.
To complete this discussion with the Maxwell equations as the main topic,
we should only note that the second set of the Maxwell equations can be
derived from the principle of least action with the Lagrangian
 $$L_{em} =
\frac{1}{2} (|E|^2 - |H|^2).$$

We also write the expression for the components
of the energy-momentum tensor in terms of electric and magnetic field
strength \be
T_{ij}= \frac{1}{2} g_{ij} (|E|^2 + |H|^2) - E_i E_j - H_i H_j -t_i P_j
-t_j P_i, \ee where $P_i$ are covariant components of the
Pointing vector $$
P^i=\varepsilon^{ijkl} t_j E_k H_l,\quad { P}=[{ E} { H}].$$

To conclude this section we mention the existence of the
Lagrangian
$$L_{CS} = \frac{q}{2} t_i A_j \tilde F^{ij}$$
that is not invariant with respect to time reversal.
The modification of electrodynamics due to the Lagrangian of this form was
considered in  [11], where an analogue of the gradient of the temporal
field was introduced to $L_{CS}$ and interpreted as a mass for the photon.
It would be very interesting to reconsider the results of the paper [11]
from the new point of view presented here.

\section{Energy}
Let us consider the energy density
\be \varepsilon = {\tilde G}_{ij}t^i t^j +
   T_{ij} t^i t^j = \varepsilon_g +\varepsilon_m.  \ee
of the full system of interacting fields.
Here $\varepsilon_g$ and $\varepsilon_m$ are the energy density of
the gravitational and material fields, respectively.
Let us write $\varepsilon_m$ for the case of the electromagnetic field.
From (53), (54) we find the energy density of the
electromagnetic field
 \be \varepsilon_{m} = \frac{1}{2} (E^2 + H^2) \ee
which is a reasonable result.

For the completeness of the picture we shall formulate the
energy conservation law starting from the general covariant Maxwell
equations.
Out of  (47), (49) we immediately get $$ E_l D_t
E^l + H_l D_t H^l  + (E^2 +H^2) \nabla_i t^i + \nabla_i P^i =0.$$
Taking into account the relations  $$ \frac{1}{2} D_t (E_l E^l) =
\frac{1}{2} D_t |E|^2 =  E_l D_t E^l + \frac{1}{2} E^l E^k D_t g_{lk},$$
$$\nabla_i t^i =\frac{1}{2} g^{ij}(\nabla_i t_j +\nabla_j t_i ) =
\frac{1}{2} g^{ij} D_t g_{ij} = \frac{1}{{\sqrt g}} D_t {\sqrt g},$$
this equation may be written in the form \be
\frac{1}{{\sqrt g}} D_t ({\sqrt g} \varepsilon_{em} ) + \nabla_i P^i = -
\frac{1}{2} T^{ij} D_t g_{ij}, \ee
where $T^{ij} = {\tilde g}^{ik} {\tilde
g}^{jl} T_{kl}.$ Let us now assume that the electromagnetic field
is considered on the background of the physical manifold of some
full system of fields. Moreover, it is known that gravity field of this
system is static, i.e., $D_t g_{ij} =0.$  In such approximation,
when physical manifold is external with respect to the electromagnetic
field, from (56) we obtain that the energy density of the
electromagnetic field satisfies the equation  $$ \frac{1}{{\sqrt g}} D_t
({\sqrt g} \varepsilon_{m} ) + \nabla_i P^i = 0. $$
It is exactly the energy conservation law of the electromagnetic
field in the above mentioned approximation.

Let us now consider the energy density of gravity
field itself in different representations.  Since
$$ \varepsilon_g  =  {\tilde G}_{ij} t^i t^j =
\frac{1}{2}{\tilde R}_{ij} g^{ij}, $$
then in the system of coordinate
compatible with causal structure for the energy density of gravity field
we have $ \varepsilon_{g} ={\tilde R}_{44} + {\tilde R}_{\mu \nu} g^{\mu
\nu}.$ From this it is not difficult to find that in this system of
coordinate energy density of gravity field can be presented in the
form \be \varepsilon_{g}=
\frac{1}{8}( (Tr K)^2 - Tr(K^2)) + \frac{1}{2} P, \ee where the matrix $K$
is defined by the relation
$$ K^{\mu}_{\nu} = g^{\mu \tau} K_{\nu\tau} = g^{\mu \tau} \partial_t
g_{\nu\tau}, $$  and $P $  is a scalar curvature of the space section of
the physical manifold of the system in question. In the  found expression,
it can be
seen that there is no term that contains second derivatives with respect to
the time coordinate. First term in the right hand side of the equation
(57) can be interpreted as the density of the kinetic energy  and second
one is the density of potential energy of the gravity field,
$ \varepsilon_g = \varepsilon_k + \varepsilon_p,  $
where
$$ \varepsilon_k = \frac{1}{8}( (Tr K)^2 - Tr(K^2)), \quad \varepsilon_p =
\frac{1}{2} P . $$

To get general covariant expression for the density of
kinetic and potential energy let us consider another representation for $
\varepsilon_{g}.$  From (29) it follows that ${\tilde R}_{ij} = R_{ij} +
2\nabla_l(t^l \nabla_i t_j).$ Thus, $$ \varepsilon_{g}= \frac{1}{2} {\tilde
R}_{ij} g^{ij} = \frac{1}{2} R + (\nabla_i t^i)^2 + D_t (\nabla_i t^i).$$
Out of the  Ricci identity $\nabla_l \nabla_k t^i - \nabla_k \nabla_l t^i =
R^i_{lkj}t^j, $ we get $t^l \nabla_l \nabla_k t^k = t^l \nabla_k \nabla_l
t^k - R_{lj} t^l t^j.$ Since $ \nabla_l t_k = \nabla_k t_l, \quad t_l t^l
=1, $ then $t^l \nabla_k \nabla_l t^k =  t^l \nabla^k \nabla_k t_l = -
\nabla^k t^l \nabla_k t_l .  $  Taking into account the relation $ 2
\nabla_k t_l = D_t g_{kl},$ for  $ \varepsilon_{g}$ we finally get \be
\varepsilon_{g}= \frac{1}{4} (g^{ik}g^{jl} - g^{ij} g^{kl}) D_t g_{ik} D_t
g_{jl} + \frac{1}{2}(R -2 R_{ij}t^it^j ).  \ee

If we now move to the system of coordinate compatible with causal
structure and compare found expression  with that established earlier, then
one can conclude that density of the kinetic energy of the gravity field
can be presented in the following general covariant form
 $$ \varepsilon_{k} =
\frac{1}{8} (g^{ik}g^{jl} - g^{ij} g^{kl}) D_t g_{ik} D_t g_{jl} ,$$
whereas density of the potential energy is given as follows
$$ \varepsilon_{p} = \frac{1}{8} (g^{ik}g^{jl}
- g^{ij} g^{kl}) D_t g_{ik} D_t g_{jl} + \frac{1}{2}(R -2 R_{ij}t^it^j ).
$$

Let us find the solution of the Einstein equations (25) for the case
when energy-momentum tensor is equal to zero and physical manifold is
flat, i.e., the Riemann tensor of the metric (1) equal to zero,
$R^i_{jkl} = 0.$  In such case physical manifold can be considered as
four-plane and hence
$g_{ij} = \delta_{ij}.$ Under such conditions equation (5) has the
following regular everywhere solution  $f(u) = a_i u^i,$ where $a_1, a_2,
a_3, a_4 $ are constants constrained by the algebraic equation
$a^2_1 + a^2_2 + a^2_3 + a^2_4 =1.$ Thus, gradient of the temporal
field is a constant unit vector  with arbitrary direction on the
four-plane. If one
considers the manifold as a background, then any physical consequences
should not depend on this arbitrariness.  Thus, the described  solution
 depends on three arbitrary parameters and it is not difficult to
verify that for this solution $ \varepsilon_{g}=0.$
Selecting one direction from the
continuum we transmute considered physical manifold into the
Minkowski space-time, which as is well-known, is a good background for the
description of the physical phenomena. However for deeper understanding
of the conceptual picture of the modern theoretical  physics we need to
develop internal field theory.

From the theory of time presented here it follows directly
that there is {\bf matter outside the time }. For example, the
gravity field and the electromagnetic field exterior to
time are described by the equations
$$ R_{ij}- \frac{1}{2} g_{ij} R = g_{ij} F^2 - F_{ik} F_{jl} g^{kl},$$
$$ \nabla_i F^{ij} = 0, \quad F^{ij} = F_{kl} g^{ik} g^{jl}, \quad F_{ij}
= \partial_i A_j - \partial_j A_i ,$$ where $F^2 = \frac{1}{4} F_{ij}
F^{ij}.$  In this context it is very important to understand the nature of
the emergence of time.

\section{Conclusion}

Let us now sum up the obtained results  and focus on some
of the problems. It is shown
that gravity physics is an internal field theory by nature which does not
contain apriori elements and can be characterized as follows. In the
theory there is no internal reason that could objectively
distinguish one arbitrary coordinate system from another.
It is unrolled on the smooth four-manifold that does not exist
apriori but is defined by the physical system itself.
Manifold is a basic primary notion in physics and representation
about space and time is given on this ground. All notions, definitions and
laws are formulated in coordinate independent form, i.e., within the
framework of the structure of smooth manifold.
Einstein's gravitational potentials are determined by the positive definite
symmetrical tensor field (Riemann metric) and temporal field.
Being a scalar field on the manifold the temporal field is
introduced into the theory in the form that provide, in general, time
reversal invariance. The system evolves in its proper time
and it is important that we need not to compare intrinsic time with
some other times. Central role of the time in the internal field
theory is that time determines causal structure of the field theory and
first of all it characterizes the internal nature of the gravity field as
the simplest closed system. It is established that energy density is the
first integral of the closed system of interacting fields.
The general covariant definition of the electric and magnetic fields is
given and the Maxwell equations for this fields are derived
( of great interest is the problem of the evolution form
of the Dirac equation on the physical manifold ).
It is shown how to reveal important internal properties
of physical system when coordinates have no physical sense. This means that
physical sense of general covariance is recognized and there exists a
reparametrization-invariant description ( see [9] for an excellent and
detailed survey on this subject).
\newpage
\begin{center} {\bf {References} \\} \end{center}

\begin{enumerate}
\item A. Einstein, Z. Math. und Phys. {\bf 62}, 225-261 (1913) (Mit
M.M. Grossmann); Naturforsch. Gesellhaft, Z\"urich, Vierteljahrschr, {\bf
58}, 284-290 (1913);  Phys. Z. {\bf 14}, 1249-1262 (1913);  Z. Math.  und
Phys. {\bf 63}, 215-225 (1913) (Mit M. M. Grossmann); Sitzungsber.  preuss.
Acad. Wiss. {\bf 48}, 844-847 (1915); Ann. Phys. {\bf 49}, 769-822 (1916).
\item  L.D. Landau and  E.M. Lifshitz, {\sl The Classical Theory of Fields
} ( Addison-Wesley and Pergamon, New York, 1971).  \item S. Weinberg, {\sl
Gravitation and Cosmology: Principles and Applications of the General
Theory of Relativity } (Wiley, New York, 1972).  \item  S.W. Hawking and
G.F.R. Ellis, {\sl The large scale structure of Space-Time} (Cambridge at
the University Press, 1973). \item C.W.  Misner, K.S. Thorne, and
J.A. Wheeler, {\sl Gravitation}  (Freeman, San Francisco, 1973).  \item
B.S. DeWitt, Phys. Rev. {\bf 160}, 1113 (1967).  \item  E. Alvarez,  Rev.
Mod. Phys.  {\bf 61}, 561 (1989).  \item  C. Rovelli, Phys. Rev. D{\bf 43},
442 (1991).  \item C. Isham, {\sl Canonical quantum gravity and the problem
of time} (Lectures presented at the NATO Advanced Study Institute,
Salamanca, June 1992); gr-qc/9210011.
\item D. Gromoll, W.Klingenberg, W.Meyer, {\sl Piemannsche Geometrie im
Grossen} ( Springer-Verlag, 1968).
\item  S.M. Carrol, G.B. Field and R. Jackiw, Phys. Rev.  D{\bf 41}, 1231
(1990).

\end{enumerate} \end{document}